\documentclass{amsart}
\usepackage{amssymb,latexsym,amsmath}
\usepackage[mathscr]{euscript}
\usepackage{lscape}
\usepackage{amsthm}
\oddsidemargin .in
\begin {document}
%%%%%%%%%%%%%%%%%%%%%%%%%%%%%%%%%%%%%%%%%%%%%%%%%%%%%%%%%%%%%%%%%%%%%%%
%%%%%%%%%%%%%%%%%%%%%%%%%     Macros      %%%%%%%%%%%%%%%%%%%%%%%%%%%%%
%%%%%%%%%%%%%%%%%%%%%%%%%%%%%%%%%%%%%%%%%%%%%%%%%%%%%%%%%%%%%%%%%%%%%%%
\numberwithin{equation}{section}
\theoremstyle{plain}% default
\newtheorem{thm}{Theorem}[section]

\theoremstyle{definition}
\newtheorem{dfn}[thm]{Definition}

% shorthand
\def\G2{{G_2}}

% blackboard style

% mathcal style

% mathfrak

% mathscript

% text style
\def\CL{\mathcal L}

% greek

% fractions

\def\ba{\begin{eqnarray}}
\def\ea{\end{eqnarray}}

\def\lb{\label}
\def\be{\begin{equation}}
\def\ee{\end{equation}}
%\def\thesection{\arabic{section}.}
%\def\thesubsection{\arabic{section}.\arabic{subsection}}
%\def\theequation{\arabic{section}.\arabic{subsection}.\arabic{equation}}
%\def\theequation{\arabic{section}.\arabic{equation}}
%%%%%%

\def\G{\Gamma}

%%%%%%%%%%%%%%%%%%%%%%%%%%%%%%%%%%%%%%%%%%%%%%%%%%%%%%%%%%%%%%%%%%%%%%%
%%%%%%%%%%%%%%%%%%%%%     Text of paper    %%%%%%%%%%%%%%%%%%%%%%%%%%%%
%%%%%%%%%%%%%%%%%%%%%%%%%%%%%%%%%%%%%%%%%%%%%%%%%%%%%%%%%%%%%%%%%%%%%%%

\title[Mirror symmetry aspects for compact ${\mathbf G_2}$ manifolds]
{Mirror symmetry aspects for compact ${\mathbf G_2}$ manifolds}
\author{Sema Salur and Osvaldo Santillan}
\date{\today}
\address {Department of Mathematics, University of Rochester, Rochester, NY,
14627.} \email{salur@math.rochester.edu }
\address {Secretaria de Cultura de Avellaneda, Buenos Aires, Argentina 1870
} \email{firenzecita@hotmail.com}
\begin{abstract}

The present paper deals with mirror symmetry aspects of compact
``barely'' $G_2$ manifolds, that is, $G_2$ manifolds of the form
(CY$\times S^1)/\mathbb{Z}_2$. We propose that the mirror of any
barely $G_2$ manifold is another barely manifold constructed as a
fibration of the \emph{mirror} of the CY base. Also, we describe the
Joyce manifolds of the first kind as ``barely'' with an underlying
CY which is self-mirror with $h^{1,1}=h^{2,1}=19$. We propose that
the mirror of a Joyce space of the first kind is another Joyce space
of the first kind. We also suggest that this self-mirror CY family
is dual to K3$\times S^1$ in the heterotic/M-theory sense. The
Borcea-Voisin construction plays a significant role for showing
this. As a spin-off we conclude that no 5-brane instantons are
present in compactifications of eleven dimensional supergravity over
Joyce manifolds of the first kind.

\end{abstract}
\maketitle
%%%%%%%%%%%%%%%%%%%%%%%%%%%%%%%%%%%%%%%%%%%%%%%%%%%%%%%%%%%%%%%%%%%%%
%%%%%%%%%%%%%%%%%%%%%%%%%%%%%%%%%%%%%%%%%%%%%%%%%%%%%%%%%%%%%%%%%%%%%
\section{Introduction}

  A {\em Calabi-Yau} manifold is a K\"{a}hler $2n$-manifold $X$ with
vanishing first Chern class and admits a Ricci-flat metric. Such
manifolds come equipped with a nowhere vanishing holomorphic
$(n,0)$-form $\Omega$ their holonomy group is $SU(n)$ or a subgroup.
Harvey and Lawson \cite{HL} showed that $Re(\Omega)$ is a
calibration on $X$. The corresponding calibrated submanifolds in $X$
are called special Lagrangian $n$-folds. The moduli space of special
Lagrangian submanifolds is expected to play a role in explaining the
mirror symmetry for Calabi-Yau (CY) manifolds.

\vspace{.1in}

CY manifolds are target spaces of (2,2) supersymmetric sigma models,
and for this reason they are physically relevant. The infinitesimal
symmetries of the models include, at classical level, two copies of
$N=2$ supersymmetry algebras without central charge. The
quantization procedure becomes very difficult for them, but is
expected to give an $(2,2)$ super-conformal quantum field theory
with central extension (or non vanishing central charge). These
theories are representations of the $(2,2)$ superconformal
algebra.
\vspace{.1in}

The mirror symmetry problem has many aspects to be covered
completely. But we can emphasize some aspects. The $(2,2)$
superconformal algebra contains the usual Virasoro generators $(L_n,
\overline{L}_n)$ and the $(2,2)$ super-symmetry generators
$(G^{\pm}_n, \overline{G}^{\pm}_n)$. It also contains two $U(1)$
currents $(J_n, \overline{J}_n)$ which are introduced to make the
algebra to close. These currents play a significant role in the
mirror symmetry problem. The transformation \be\lb{auto} L_n
\longrightarrow L_n, \qquad J_n\longrightarrow -J_n, \qquad
G^{\pm}_n\longrightarrow G^{\pm}_n, \ee is an automorphism of the
$(2,2)$ algebra and is called left mirror automorphism. The
analogous definition hold for right mirror automorphism, and when it
affects to both sectors it will be a target space automorphism. The
transformation (\ref{auto}) just reverse the sign of the $U(1)$
charge of a given state corresponding to $J$.
\vspace{.1in}

 The mirror automorphism has several important consequences.
An important set of operators of $(2,2)$ superconformal theories are
are marginal operators, which are the ones for which the conformal
weight sums $h+\overline{h}=2$. These operators can be used to
deform a given conformal theory to a nearly conformal field theory
without changing the central charge. But in order to obtain a
continuous family of conformal field theories one should consider
truly marginal operators, that is, operators which are still
marginal after deformation. It follows that the effect of these
operators is to induce deformations of a CY preserving the CY
property. These deformations are known to come in two types:
deformations of the K\"{a}hler structure and of the complex
structure, and are captured by cohomology groups $h^{1,1}$ and
$h^{2,1}$ for 3-folds. This is a beautiful link between abstract
conformal field theory aspects and the geometry of the target
manifold \cite{Witten}.
\vspace{.1in}

In fact, the two different types of marginal operators differ only
in the sign of the $U(1)$ charge. This means that the automorphism
generate a transformation \be\lb{mirrsim} h^{1,1}\leftrightarrow
h^{2,1} \ee between the Hodge numbers of the CY. In other words, it
corresponds to a change in the Euler signature $\chi=2(h^{1,1} -
h^{2,1})$ and therefore it result in a topologically different CY
manifold. This is uncomfortable asymmetry which implies that is not
possible to reconstruct the target space geometry from sigma model
data. In order to avoid it it was postulated that CY manifolds come
in pairs with the Hodge numbers related by (\ref{mirrsim}) such that
the resulting superconformal theories are isomorphic. Such manifolds
will be called mirror to each other.
\vspace{.1in}

   After this hypothesis was introduced several examples satisfying
(\ref{mirrsim}) were constructed in \cite{Candelas}. Nevertheless
(\ref{mirrsim}) is a necessary condition for two manifolds to form
mirror pairs, but is not sufficient. Two manifolds are truly mirror
pairs if they correspond to the \emph{isomorphic} conformal field
theories and these result in a relation involving 3-point
correlation functions from both sides. Apparently the first explicit
realization of mirror pairs were found in \cite{Plesser}. The idea
was to find a group of automorphism of the conformal theory which is
not an automorphism of the underlying CY space. This automorphism
will automatically generate a mirror pair.
\vspace{.1in}

   An alternative definition of mirror symmetry comes from Kaluza-Klein
reductions. If one consider, for instance, compactification of IIA
superstings on a background of the form $M_4\times K_6$ to
4-dimensions, then the condition of $N=1$ supersymmetry for the low
dimensional theory implies that $K_6$ is Calabi-Yau. The change of
sign of the $U(1)$ charge reverse the GSO projection. This situation
is analogous of a T-duality transformation on a two torus in which
IIA superstring is mapped to IIB and viceversa. Therefore if IIA and
IIB theories are compactified to $D=4$ over two mirror Calabi-Yau
three-folds then the resulting 4-dimensional theories are expected
to be isomorphic.
\vspace{.1in}

An interesting question is if any CY manifold has a mirror and if
the mirror transformation is somehow related to T-duality. The
analysis of II compactifications including instanton corrections
lead to the SYZ conjecture which states that if a CY three fold has
a mirror, then both manifolds are locally a special Lagrangian
$\mathbb{T}^3$ fibration and are related to each other by a
T-duality acting on each coordinate of the $\mathbb{T}^3$
\cite{Zaslow}. But a complete proof of this conjecture is still
lacking.

\vspace{.1in}

The mirror symmetry conjecture was subsequently generalized to other
type of manifolds. The ``generalized'' mirror conjecture
\cite{ShataVafa} states that if there is an ambiguity in determining
the topological properties of the target manifold in a sigma model,
then there exists a dual manifold resolving the ambiguity. Another
possible definition is that a pair of manifolds $(X, \overline{X})$
is a mirror pair if compactifying IIA and IIB supergravities over
them gives isomorphic low dimensional theories \cite{Acharya3}. The
natural problem is to understand how the topological invariants of
mirror manifolds are related, i.e, to find an analog of the relation
(\ref{mirrsim}) for any kind of mirror set.

\vspace{.1in}

The present paper deals with mirror symmetry aspects of \emph{$G_2$
holonomy manifolds}\footnote{A more mathematical discussion about
this subject can be found in the recent paper \cite{akbulut}.}.
These are 7-dimensional and are characterized by two $G_2$
equivariant 3-form $\varphi$ and 4-form $*\varphi$ which are both
closed. As CY manifolds, these are Ricci flat and Harvey and Lawson,
\cite{HL} showed that $\varphi$ and $*\varphi$ are calibrations on
$M$. The corresponding calibrated sub-manifolds in $M$ are called
associative $3$-folds and co-associative $4$-folds, respectively. We
will focus on \emph{compact} $G_2$ holonomy manifolds. Compactness
is required for Kaluza-Klein matters, in order to obtain a discrete
Kaluza-Klein spectrum.

\vspace{.1in}

   The sigma model analysis of mirror $G_2$ manifolds was performed in
\cite{ShataVafa}-\cite{Howe}. There was shown that if one consider a
sigma model with $N=1$ supersymmetry over a $G_2$ manifold the
effect of the calibrations $\varphi$ and $*\varphi$ is to add two
new operators $\Phi$ and $X$ with spin 3/2 and 2 to the N=1
generators $T$ and $G$, and also two more operators $K$ and $M$ of
spin 2 and 5/2 in order the algebra close. This is called extended
supersymmetry algebra. But the important thing is that $\Phi$ and
$X$ generate themselves a new N=1 superconformal sub-algebra with
central charge 7/10, which corresponds to the tri-critical Ising
minimal model. By use of this, the authors of \cite{ShataVafa} were
able to classify the highest weight states of the algebra in terms
of the tri-critical Ising highest weight and the eigenvalue of the
remaining stress energy tensor. Also, they identified the marginal
deformations of the theory and showed that the physical moduli space
has dimension $b_3+b_2$. This is different than dimension
geometrical moduli space dimension for $G_2$ manifolds, which is
$b_3$. This discrepancy is due to the physical freedom to add a
closed two form to add a phase to the action, which has no geometric
analog.

\vspace{.1in}

In fact, the physical moduli space dimension $b_3+b_2$ is in
agreement with the results of \cite{PapadopoulosTownsend}, where it
was shown that IIA and IIB compactifications over the same compact
$G_2$ holonomy space give "apparently" the same field theory
content, namely $b_2+1$ scalar multiplets and $b_3$ vector
multiplets. Guided by this fact, the authors of
\cite{PapadopoulosTownsend} raised the possibility that both
compactifications are equivalent, in other words, that $G_2$
manifolds are mirror to themselves. Some examples realizing these
were suggested in \cite{Acharya2}, but also a counterexample. But
there is a subtlety in three dimensions, which is a duality
transformation taking scalar into vectors and vice versa. Therefore
the analysis of \cite{PapadopoulosTownsend} does not collect
manifolds with the same $b_2$ and $b_3$, but instead those with the
same value of $b_2+b_3$, as predicted by \cite{ShataVafa}.

\vspace{.1in}

Besides these developments, the problem of mirror symmetry for $G_2$
manifolds is less understood than for the CY. Two manifolds with the
same $b_2+b_3$ value are not necessarily mirror, the condition to
give isomorphic physics should stronger than that, just as in the CY
case. An interesting analysis of mirror $G_2$ manifolds was made in
\cite{Acharya3}, where it was suggested that $G_2$ manifolds
admitting a mirror should possess four cycles $C_4$ which satisfy
the condition $b_2^+(C_4) + b_3(C_4)=7$. The only known example is
the four torus $T^4$ and therefore this suggest that mirror pairs
are locally $T^4$ fibrations. This statement will be the analogous
of the SYZ conjecture for $G_2$ manifolds. But its validity relies
in our ability to prove that there are no other solutions $C_4$ of
the equation $b_2^+(C_4) + b_3(C_4)=7$, and this is still an open
question.

\vspace{.1in}

  An important discovery is also the topological side
for both $(2,2)$ superconformal models over CY and N=1 sigma models
over $G_2$ manifolds. In \cite{Verlinde} the authors considered
conformal field theories for which the stress tensor $T_{\mu\nu}$ is
BRST-trivial, i.e, is of the form $T_{\mu\nu}=\{Q, G_{\mu\nu}\}$,
being $Q$ a nilpotent operator $Q^2=0$. This theories will be
automatically topological and the topological conformal algebra
between the operators $T$, $G$ and $Q$ was worked out explicitly
\cite{Verlinde}. The result is an conformal algebra without central
charge which is related to the N=2 superconformal algebra by a
redefinition of the stress energy tensor as $T \to
T+\frac{1}{2}\partial J$. This redefinition is called a twist. In
similar lines, the authors of \cite{ShataVafa} found a redefinition
of the operators $X$ and which has the effect of switch the central
charge to zero. This remarkable result is an strong hint for the
existence of a topolgical field theory associated to $G_2$
manifolds, which was worked out in \cite{Boer}-\cite{Boer2} in more
detail.

\vspace{.1in}

    Another relevant branch is
the presence of associative and coassociative submanifolds, which
are the calibrated submanifolds of $G_2$ holonomy manifolds . As is
well known, eleven dimensional supergravity compactified in a $G_2$
manifold gives $N=1$ supersymmetry in the low effective theory. But
supergravity also contains solitons breaking the supersymmetries of
the theory, unless certain restrictions are satisfied \cite{Becker}.
These restrictions have been found to be equivalent to the presence
of the calibrated submanifolds inside the compactification $G_2$
space. In other words, calibrated submanifolds of $G_2$ manifolds
are supersymmetric cycles \cite{Becker}.

\vspace{.1in}

The present paper is organized as follows. In \S\ref{msa2}, we
describe the K3 surfaces together with equivalent heterotic/M-theory
compactifications over K3$\times S^1$/CY spaces and identify the CY
manifold involved in this duality. This CY turns to be important in
our further discussions. In \S\ref{msa3} we discuss general aspects
of compact $G_2$ manifolds, in particular calibrated submanifolds
and almost Calabi-Yau structures inside them. In \S\ref{msa4}, we
review the construction of Joyce manifolds of the first kind and we
present some aspects of M-theory/heterotic string dualities over
$G_2$/CY manifolds and also of mirror symmetry for $G_2$ manifolds.
We propose that the evidence found in \cite{PapadopoulosTownsend}
that the mirror map leave inert the betti numbers of the Joyce
manifold is because they are fibrations over CY with zero Euler
number, which are in some sense "protected" from topology change
under the complex/sympletic map. We generalize the discussion to
general "barely" $G_2$ manifolds. Also, we show that for
compactification of Joyce spaces of the first kind no 5-brane
instanton appears.

\section{Compact self-dual metrics on K3}

\label{msa2}

    K3 surfaces play an important role
in several string dualities and in the Joyce construction of compact
$G_2$ holonomy manifolds. For these reasons, we discuss them first,
together with the mentioned dualities.

\subsection{Hyperk\"{a}hler metrics over K3}

  As we stated in the introduction, Calabi-Yau spaces
are 2m-dimensional with holonomy in $SU(m)$. By another side,
hyperk\"{a}hler spaces are by definition 4n-dimensional with
holonomy in $Sp(n)$. Both spaces admit a Ricci-flat K\"{a}hler
metric. The dimension $D=4$ is special because it corresponds to
$n=1$ and $m=2$ and by the isomorphism $Sp(1)\simeq SU(2)$, it
provides a link between both cases. If the manifold is compact and
$c_1=0$, then the Yau proof of the Calabi conjecture imply that it
admits a unique Ricci flat K\"{a}hler metric \cite{Calabi}. In $D=4$
this metric will be simultaneously Calabi-Yau and hyperk\"{a}hler.
\vspace{.1in}

The curvature of the hyperkahler 4-metrics is always self-dual
$$
R_{abcd}=\ast R_{abcd}=\frac{1}{2}\epsilon_{abef}R^{ef}_{cd},
$$
and this implies Ricci-flatness, i.e, $R_{ij}=0$ where $R_{ij}$ is
the Ricci tensor of the metric. Besides, there exists a vielbein
basis $e^i$ in which they are expressed as
$g_4=\delta_{ij}e^i\otimes e^j$ and for which the hyperk\"{a}hler
triplet \be\lb{triplet} \overline{J}_1=e^1\wedge e^2+e^3\wedge e^4,
\qquad \overline{J}_2=e^1\wedge e^3+e^4\wedge e^2, \qquad
\overline{J}_3=e^1\wedge e^4+e^2\wedge e^3 \ee is closed.
\vspace{.1in}

    In dimension four, several explicit hyperk\"{a}hler metrics are known,
but they are all defined over \emph{non-compact} manifolds.  One of
the best known examples is the Eguchi-Hanson gravitational instanton
\cite{Eguchi-Hanson}, which possess a complete hyperk\"{a}hler
metric given by
\begin{equation}
g=\frac{r^{2}}{4}\left( \;1-(a/r)^{4}\;\right) \;(\;d\theta +\cos
\varphi
d\tau \;)^{2}+\left( \;1-(a/r)^{4}\;\right) ^{-1}\;dr^{2}+\frac{r^{2}}{4}%
\;(\;d\varphi ^{2}+\sin ^{2}\varphi d\tau^2 \;).  \label{ego}
\end{equation}
This metric contains an $S^2$ sphere of radius $a$ at its tip. If
the parameter $a$ tends to zero the result will be the
hyperk\"{a}hler flat metric on $\mathbb {C}^2/\mathbb{Z}_2$. In
fact, the Eguchi-Hanson space is an ALE (asymptotically Euclidean
space), which means that it approaches asymptotically to the
Euclidean metric. The boundary at infinity is locally $S^{3}$.
However, the situation is rather different in what regards its
global properties. This can be seen by defining the new coordinate
\vspace{.1in}
\[u^{2}=r^{2}\left( 1-(a/r)^{4}\right)\]%
\vspace{.1in} \noindent for which the metric can be rewritten as
\begin{equation}
g=\frac{u^{2}}{4}\;(\;d\theta +\cos \varphi d\tau \;)^{2}+\left(
\;1+(a/r)^{4}\;\right) ^{-2}\;du^{2}+\frac{r^{2}}{4}\;(\;d\varphi
^{2}+\sin ^{2}\varphi d\tau \;).  \label{ego2}
\end{equation}%
The apparent singularity at $r=a$ has been moved now to $u=0$. Near
the singularity, the metric looks like
$$
g\simeq \frac{u^{2}}{4}\;(\;d\theta +\cos \varphi d\tau
\;)^{2}+\frac{1}{4}%
du^{2}+\frac{a^{2}}{4}\;(\;d\varphi ^{2}+\sin ^{2}\varphi d\tau \;),
$$
and, at fixed $\tau $ and $\varphi ,$ it becomes
$$
g\simeq \frac{u^{2}}{4}\;d\theta ^{2}+\frac{1}{4}du^{2}.
$$
This expression ``locally'' looks like the removable singularity of
${\mathbb R}^{2}$ that appears in polar coordinates. However, for
actual polar coordinates, the range of $\theta $ covers from $0$ to
$2\pi $, while in spherical coordinates in ${\mathbb R}^{4},$ $0\leq
\theta <4\pi $. This means that the opposite points on the geometry
turn out to be identified and thus the boundary at infinity is the
lens space $S^{3}/{\mathbb{Z}}_{2}$, which is the same boundary as
for $B_4/{\mathbb Z}_2$.
 \vspace{.1in}

  To construct \emph{compact} hyperkahler manifolds is more complicated, but
there is an ingenious way to prove their existence \cite{Page},
\cite{Lebrun}. The Eguchi-Hanson instanton plays a significant role
in this proof. Consider a 4-torus ${\mathbb T}^4={\mathbb
R}^4/{\mathbb Z}_4$, where ${\mathbb Z}_4$ is generated by the
canonical four lattice. Choose its flat metric and the
hyperk\"{a}hler triplet \be\lb{triplet} \overline{J}_1=dx^1\wedge
dx^2+dx^3\wedge dx^4, \qquad \overline{J}_2=dx^1\wedge
dx^3+dx^4\wedge dx^2, \qquad \overline{J}_3=dx^1\wedge
dx^4+dx^2\wedge dx^3, \ee and identify the points of this torus by a
${\mathbb Z}_2$ action which reflects through the origin. This
action preserves the triplet (\ref{triplet}) and has the fixed
points $(r_1, r_2, r_3, r_4)$ with $r_i$ taking values $0$ or $1/2$.
This gives a total of $4\times 4= 16$ fixed points. The
corresponding singularities are of type $A_1$ and the holonomy of
the resulting space is ${\mathbb Z_2}$. The Kummer construction of a
compact hyperk\"{a}hler metric consist in excising a region of
radius R around all the $A_1$ singularities, which gives the
topology of a ball $B_4/{\mathbb Z}_2$, and replacing it by a copy
of a Eguchi-Hanson space. The modified metric around a singularity
will be given by the expression (\ref{ego}) with the modification
$a\to a\; \tau(|\overrightarrow{R}-\overrightarrow{r}_i|)$ where
$\tau (x)$ is a smooth function which takes value 1 in the region
$x\leq R_1$ and zero for $x> R_2$. Here $\overrightarrow{R}$ is a
point in the manifold and $\overrightarrow{r}_i$ is the position of
an i-th singularity. As a result it will be a smooth manifold $M$
and we have a map $\pi: M\to {\mathbb T}^4/{\mathbb Z}_2$ which is
called the resolving map.

\vspace{.1in}

   The reason for choosing the Eguchi-Hanson space is justified by its
topological properties: it is an asymptotically flat self-dual
metric with a natural ${\mathbb Z}_2$ action which matches the
${\mathbb Z}_2$ action on ${\mathbb T}^4/{\mathbb Z}_2$. Thus,
although the imperfect matching in the boundary of $B_4/{\mathbb
Z}_2$, the resulting metric is an approximation for a compact
hyperk\"{a}hler metric. It can be shown that the failure of the
triplet (\ref{triplet}) to be closed is of the order of
$O(a^4/R^4)$. Clearly, we can be as closer as we want to a
hyperk\"{a}hler metric by taking a sufficiently small value of $a$,
but the limit $a \to 0$ will give our initial orbifold. However for
small enough values of $a$ it is possible to deform it to an smooth
structure $(M, g, \overline{J}_{i})$ with holonomy \emph{exactly}
$SU(2)$, \cite{Lebrun}. This procedure is called a blowup, and it
relies on deformation theory of singular complex manifolds. Thus
compact hyperk\"{a}hler metrics do exist, although nobody has found
their explicit form.

\vspace{.1in}

   It can be proved that for the compact metrics described
above $b_1=c_1=0$. As they are obviously K\"{a}hler, they are
complex. Regular complex and compact surfaces are called K3
surfaces. It was proved by Kodaira that every K3 surface is a
deformation of a non-singular quartic Kummer surface \be\lb{homo}
f=x_1^4+x_2^4+x_3^4+x_4^4=0 \ee in $\mathbb {C}P^3$ with homogeneous
coordinates $(x_1, x_2, x_3, x_4)$, \cite{Kodaira}. Hence K3
surfaces are K\"{a}hler and all diffeomorphic and the Yau proof
implies that they admit an unique Ricci flat K\"{a}hler metric. This
can be only the metric we described in this section. \vspace{.1in}

\subsection{Equivalent compactifications related to K3 surfaces}

   It is well known that eleven
dimensional supergravity compactified on a circle $S^1$ gives the
strong coupling limit of IIA theory. But if this theory is
compactified on $S^1/{\mathbb Z}_2$ then the ${\mathbb Z}_2$ action
takes out one of the two supersymmetries of the 10-dimensional
theory. The two fixed points are two 10-dimensional hyperplanes. The
anomaly cancellation on them requires the gauge group to be
$E_8\times E_8$, which arises from twisted sectors of the orbifold.
It has been suggested that the resulting theory is heterotic string
theory on $E_8\times E_8$ \cite{Horava}.

\vspace{.1in}

  The low energy limit of M-theory is eleven dimensional
supergravity. If we go to seven dimensions by Kaluza-Klein reduction
over ${\mathbb T}^4$ the supersymmetries will be maximal.
Compactifications over the orbifold ${\mathbb T}^4/{\mathbb Z}_2$
break half of them and the same will occur in the K3 limit described
in the previous section. It has been shown that the result of the
compactification is an Einstein-Maxwell supergravity theory $D=7$
with a three form coupled to the $D=7$ supermembrane. There is
evidence showing that this theory is the effective action of
heterotic string on ${\mathbb T}^3$ at strong coupling. Thus we have
heterotic/M-theory equivalence over ${\mathbb T}^3/K3$. Also IIA
compactifications over $K_3\times {\mathbb T}^2$ have been
conjectured to be equivalent to heterotic string theory over
${\mathbb T}^6$ \cite{Hulltown}.

\vspace{.1in}

    In view of these equivalences, it is natural to consider $K3\times
S^1$ compactifications of heterotic string and to investigate if
there is a compactification from eleven dimensions to five giving
the same theory \cite{PapadopoulosTownsend}. The compactification
manifolds should be six dimensional and the task is to identify it.
From the heterotic side it is obtained $N=2$ supergravity in five
dimensions coupled to 18 vector multiplets and 20 hypermultiplets.
In order to obtain $N=2$ supersymmetry from eleven dimensions, the
internal space should be CY. After compactification over the CY it
is obtained $h^{1,1}-1$ vector multiplets and $h^{2,1}+1$
hypermultiplets, where $h^{1,1}$ and $h^{2,1}$ are the Hodge numbers
of the unknown Calabi-Yau. Therefore the number of multiplets
matches if and only if $h^{1,1}-1=18$ and $h^{2,1}+1=20$, in other
words, if \be\lb{papaton} h^{1,1}=h^{2,1}=19. \ee The constraint
(\ref{papaton}) implies that the Hodge numbers of the internal
manifold are invariant under the mirror transformation
(\ref{mirrsim}), and that its Euler number $\chi=2(h^{1,1}-h^{2,1})$
vanishes.

\vspace{.1in}

    In fact, there exists a large family of CY constructed as K3 fibrations
\cite{Borcea}, \cite{Voisin} which include examples satisfying
(\ref{papaton}). To construct them one, consider a K3 surface $X$
with an holomorphic 2-form $\overline{J}$ together with an action
$\sigma$ which acts by $\sigma^{\ast}(\overline{J})=-\overline{J}$.
The action $\sigma$  has a set of fixed points $\Sigma$ which has
been classified by Nikulin \cite{Nikulin}. This classification
states that $\Sigma$ is a disjoint union of smooth curves in $X$ and
there are three possibilities: either $\Sigma$ is empty, or
$\Sigma=C_1\cup C_1^{\prime}$ where $C_1$ and $C_1^{\prime}$ are
both elliptic curves or $\Sigma=C_g+E_1+\,\cdots\,+E_k$ where $C_g$
is a curve of genus $g$ and $E_i$ are rational curves.

\vspace{.1in}

  Now consider an elliptic curve $E$ and an involution (-1) which changes
the sign of its coordinates, then the 6 dimensional quotient
\be\lb{borcea} X_6=\frac{K_3\times E}{(\sigma, -1)} \ee possess a
fixed point set $S_0$ consisting in four copies of $\Sigma$ and the
quotient $(X\times E)/(\sigma,-1)$ has $A_1$-singularities along
$S_0$ \cite{Borcea}, \cite{Voisin}. These singularities can be blown
up in order to give a smooth CY manifold, as for K3 manifolds. The
resulting CY´s are called Borcea-Voisin 3-folds and their Hodge
numbers are given by \be\lb{mu} h^{1,1}(Y)\;=\;11+5n-n^{\prime},
\qquad h^{2,1}(Y)\;=\;11+5n^{\prime}-n,\ee where $n$ is the number
of components of the singular set $\Sigma$ of a the K3 surface and
$n^{\prime}$ is the sum of the genus of all these components.

\vspace{.1in}

    In principle the Borcea-Voisin family (\ref{borcea}) is very large,
but if one is looking for an specific CY with the Hodge numbers
given in (\ref{papaton}) then $n=n^{\prime}=2$. This means that
$\Sigma$ should composed by two components of genus 1. Such
components should be two 2-torus. Although this is not enough to
determine the elliptic curve $E$, this deduction shows that CY
manifolds satisfying (\ref{papaton}) exist and is reasonable to
suppose that they are included in this subfamily. In fact K3$\times
S^1$ arise by smoothing the orbifold ${\mathbb T}^4/{\mathbb Z}_2
\times S^1$. Compactification of heterotic string over such orbifold
will give the same number of supersymmetries than compactifications
over the orbifold $({\mathbb T}^6/{\mathbb Z}_2\otimes {\mathbb
Z}_2)$ if the two ${\mathbb Z}_2$ acts over the whole ${\mathbb
T}^6$. All these facts suggest the following:

\vspace{.1in}

{\bf Conjecture} The CY dual to $K3$ in the heterotic/M-theory sense
is obtained as a quotient of the form (\ref{borcea}) in which the
elliptic curve $E$ is a 2-torus and the action $\sigma$ over $K3$
has a singular set which is the union of two 2-tori.

\vspace{.1in}

   If the statement above is correct, then we are in the second case in
the Nikulin classification. In fact, the dual of this
compactification has also been considered in \cite{Acharya1} and, as
far as we understand, our conjecture is in agreement with that
reference.

\section{${\mathbf G_2}$ Manifolds}

\label{msa3}

\subsection{Generalities}

 We now turn our attention to $G_2$ holonomy manifolds. We review
their basic properties, the reader can find more information in
Harvey and Lawson, \cite{HL}.

\vspace{.1in}

The group $G_2$ can be considered as the group of automorphisms of
the imaginary octonions. The octonions $\mathbb{O}=\mathbb{H}\oplus
l \mathbb{H}=\mathbb {R}^8$ constitute an 8-dimensional division
algebra and are obtained from the quaternions $\mathbb H$ using the
Cayley-Dickson process. This algebra is generated by $<1, i, j, k,
l, li ,lj, lk> $, where $i$, $j$, $k$ are the pure quaternion units.
The imaginary octonions Im$ \mathbb{O} =\mathbb {R}^7$ are naturally
equipped with the cross product operation $\times: \mathbb
{R}^7\times \mathbb {R}^7 \to \mathbb {R}^7$ defined by $u\times
v=im(u.\bar{v})$. Then one can define the exceptional Lie group
$G_{2}$ as the linear automorphisms of Im$\mathbb{O}$ preserving
this cross product operation, i.e, $u\times v=g u\times gv$ if $g\in
G_2$. Alternatively, $G_2$ is the subgroup of $GL(7,\mathbb {R})$
which fixes a particular $3$-form $\varphi_{0} \in
\Omega^{3}(\mathbb {R} ^{7})$ given below. Denote
$e^{ijk}=dx^{i}\wedge dx^{j} \wedge dx^{k}\in \Omega^{3}(\mathbb
{R}^7)$, then
$$ G_{2}=\{ A \in GL(7,\mathbb {R}) \; | \; A^{*} \varphi_{0} =\varphi_{0}\;
\}. $$
%\vspace{0.1in}
\begin{equation}\lb{octon}
\varphi_0=e^{127} +e^{136}+e^{145} +e^{235}+e^{426} +e^{347}
+e^{567} ,
\end{equation}

\vspace{.1in}

\begin{dfn}
A smooth $7$-manifold $M^7$ has a {\it $G_{2}$ structure} if its
tangent frame bundle reduces to a $G_{2}$ bundle. Equivalently,
$M^7$ has a {\it $G_{2}$ structure} if there is  a 3-form $\varphi
\in \Omega^{3}(M)$  such that  at each $x\in  M$ the pair $
(T_{x}(M), \varphi (x) )$ is  isomorphic to $(T_{0}( \mathbb
{R}^{7}), \varphi_{0})$. We call $(M,\varphi)$ a manifold with $G_2$
structure.
\end{dfn}

\vspace{.1in}

A $G_{2}$ structure $\varphi$ on $M^7$ gives an orientation $\mu \in
\Omega^{7}(M)$ on $M$, and $\mu$ determines a metric $g=g_{\varphi
}= \langle \;,\;\rangle$ on $M$, and a cross product structure
$\times$  on the tangent bundle of $M$ defined as

\begin{equation*}
\langle u,v \rangle=[ i_{u}(\varphi ) \wedge i_{v}(\varphi )\wedge
\varphi  ]/\mu .
\end{equation*}
\begin{equation*}
\varphi (u,v,w) = \langle u\times v,w \rangle .
\end{equation*}
\noindent where $i_{v}=v\lrcorner $ be the interior product with a
vector $v$.
\begin{dfn}

A manifold with $G_{2}$ structure $(M,\varphi)$  is called a {\it
$G_{2}$ manifold} if the holonomy group of the Levi-Civita
connection (of the metric $g_{\varphi }$) is a subgroup of  $G_2$.
Equivalently, $(M,\varphi)$ is a $G_{2}$ manifold if $\varphi $ is
parallel with respect to the metric $g_{\varphi }$,
$\nabla_{g_{\varphi }}(\varphi)=0$; which is equivalent to
$d\varphi=0 $, $\;d(*_{g_{\varphi}}\varphi)=0$. This implies that at
each point $x_{0}\in M$ there is a chart $(U,x_{0}) \to (\mathbb
{R}^{7},0)$ on which $\varphi $ equals to $\varphi_{0}$ up to second
order term, i.e. on the image of $U$ $\varphi (x)=\varphi_{0} +
O(|x|^2)$.

\end{dfn}

\vspace{.1in}

   We can paraphrase this definition by saying that for $G_2$ holonomy
manifolds there exist locally a 7-vein basis such that
$g_7=\delta_{ab}e^a \otimes e^b$ and for which the 3-form
(\ref{octon}) and its dual are closed, being now $e^{ijk}=e^i\wedge
e^j\wedge e^k$.

\vspace{.1in}

\subsection{Calibrated submanifolds}

The reduction of the holonomy of a given manifold from $SO(7)$ to
$G_2$ implies the existence of a covariantly constant spinor $\eta$,
i.e, an spinor satisfying $D\eta=0$ where $D$ is the spin connection
of the seven manifold. This is a well known feature in Kaluza-Klein
compactifications. For 11-supergravity compactified over a
7-manifold, the numbers of supersymmetries of the resulting
4-dimensional theory is equal to the number of Killing spinor of the
internal manifold. In particular, if the holonomy of the 7-manifold
is $G_2$, then the number of supersymmetries is one. But eleven
dimensional supergravity contains membrane solitons. These solitons
break the supersymmetries of the theory except if certain
restrictions are satisfied \cite{Becker}. These have been found to
be equivalent to the presence of calibrated submanifolds inside the
compactification $G_2$ space. There are two type of calibrated
submanifolds in a $G_2$ holonomy manifold namely, associative and
coassociative submanifolds.

\vspace{.1in}

\begin{dfn}
Let $(M, \varphi )$ be a $G_2$ manifold. A 4-dimensional submanifold
$X\subset M$ is called {\em coassociative } if $\varphi|_X=0$. A
3-dimensional submanifold $Y\subset M$ is called {\em associative}
if $\varphi|_Y\equiv vol(Y)$; this condition is equivalent to
$\chi|_Y\equiv 0$,  where $\chi \in \Omega^{3}(M, TM)$ is the
tangent bundle valued 3-form defined by the identity:
\begin{equation*}\label{chil}
\langle \chi (u,v,w) , z \rangle=*\varphi  (u,v,w,z)
\end{equation*}
\end{dfn}

\vspace{.1in}

As they solve the conditions for unbroken symmetry of \cite{Becker},
associative and coassociative submanifolds are sometimes called
supersymmetric 3- or 4-cycles, respectively. D-branes wrapping these
cycles will be supersymmetric.

\vspace{.1in}

\subsection{Relation with CY manifolds}

  The mirror map for CY relate deformations of the complex structure
of one manifold to deformations of the sympletic structure of the
mirror and viceversa. The question is how the mirror map acts on
$G_2$ manifolds. It could be interesting to consider $G_2$ manifolds
which are fibrations over CY three-folds, and to see how the mirror
map on the internal CY affect the entire 7-manifold. Some examples
of this situation are the barely $G_2$ manifolds, which we will
consider in the following sections.

\vspace{.1in}

  For a $G_2$ manifold it was shown in \cite{Salur} that,
similar to the definition (\ref{chil}) of $\chi$, one can also
define a tangent bundle 2-form $\psi$, which is just the cross
product of $M$.
\begin{dfn}
Let $(M, \varphi )$ be a $G_2$ manifold. Then $\psi \in
\Omega^{2}(M, TM)$ is the tangent bundle valued 2-form defined by
the identity:
\begin{equation*}
\langle \psi (u,v) , w \rangle=\varphi  (u,v,w)=\langle u\times v ,
w \rangle
\end{equation*}
\end{dfn}
Also, let $(M^7, \varphi , \Lambda)$ be a $G_2$ manifold with a
non-vanishing oriented $2$-plane field $\Lambda$. One can view
$(M^7, \varphi)$ as an analog of a symplectic manifold, and the
$2$-plane field $\Lambda $ as an analog of a complex structure
taming $\varphi$. This is possible because $\Lambda $ along with
$\varphi $ gives the associative/complex bundle splitting $T(M)={\bf
E}_{\varphi,\Lambda}\oplus {\bf V}_{\varphi,\Lambda}$. Now, a choice
of a non-vanishing unit vector field $\xi \in \Omega^{0}(M,TM)$,
gives a codimension one distribution $V_{\xi}:= \xi^{\perp}$ on $M$
with interesting structures induced from $\varphi$.
\vspace{.1in}

\begin{dfn}
$(X^6,  \omega, Re\;\Omega, J)$  is called an {\it almost Calabi-Yau
manifold}, if $X$ is a Riemannian manifold with a non-degenerate
$2$-form $\omega$ (i.e. $\omega^3 =vol (X)$) which is co-closed, and
$J$ is a metric invariant almost complex structure which is
compatible with $\omega$, and $Re\;\Omega $ is a closed
non-vanishing $3$ form. Furthermore, when $J$ is integrable,
$\omega$ is closed and $ Re\; \Omega$ is co-closed we call this a
Calabi-Yau manifold.
\end{dfn}

\vspace{.1in}

Then the following theorem can be proved \cite{Salur}.
\vspace{.1in}

\begin{thm} \cite{Salur}, \label{thm1}
Let $(M,\varphi)$ be a $G_2$ manifold, and $\xi $ be a unit vector
field which comes  from a codimension one foliation on $M$, then
$(X_{\xi},\omega_{\xi}, \Omega_{\xi},J_{\xi})$ is an almost
Calabi-Yau manifold with $\varphi |_{X_{\xi}}= Re\; \Omega_{\xi} $
and  $*\varphi |_{X_{\xi}}= \star \omega_{\xi} $. Furthermore, if
$\CL_{\xi}(\varphi )|_{X_{\xi}}=0$ then $d\omega_{\xi}=0$,  and if
$\CL_{\xi}(*\varphi)|_{X_{\xi}}=0$ then $J_{\xi}$ is integrable;
when both of these conditions are satisfied then
$(X_{\xi},\omega_{\xi}, \Omega_{\xi},J_{\xi})$  is a Calabi-Yau
manifold.
\end{thm}

\vspace{.1in}

The main idea behind the formalism presented above is to use $\chi$
and $\psi$ on the $G_2$ manifold to obtain the complex and
symplectic strutures on CY manifolds inside the $G_2$ manifold.
Theorem \ref{thm1} implies that both complex and symplectic
structure of the CY-manifold $X_{\xi}$ are determined by $\varphi$.
Moreover, the choice of $\xi$ can give rise to very different
complex structures on $X_{\xi}$ (i.e. $SU(2)$ and $SU(3)$
structures). So if we assume that $\xi \in \Omega^{0} (M, {\bf V})$
and $\xi'\in \Omega^{0} (M, {\bf E})$ are two unit vector fields,
and let $X_{\xi }$ and $X_{\xi '}$ are pages of the corresponding
codimension one foliations then using theorem \ref{thm1}, we showed
that one can obtain two CY manifolds with different complex
structures and called them ``dual'' in that sense.

\vspace{.1in}

\subsection{A simple example: $\mathbb {T}^7$}

As an application of the previous notions let us consider the
simplest compact $G_2$ holonomy manifold, namely $\mathbb{T}^7$,
\cite{Salur}. Although this is a trivial example, several compact
manifolds with $G_2$ are obtaining by orbifolds of the 7-torus by
perturbing them to smooth $G_2$ holonomy metrics. We expect that
many features of this example are preserved after perturbing to a
smooth $G_2$ metric. So, let us consider the calibration 3-form
(\ref{octon}) for $\mathbb{T}^7$, which is given as
\begin{equation}\lb{octon1}
\varphi=e^{127} +e^{136}+e^{145} +e^{235}+e^{426} +e^{347} +e^{567}.
\end{equation}

Note that this form is different than the 3-form $\varphi$ used in
\cite{Salur}. The reason to use other coordinates is to make our
calculations compatible with the ones in \cite{Liu}. From
\cite{Salur}, we have the decomposition $T(M)={\bf E}\oplus {\bf
V}$, where ${\bf E}=\{e_1,e_2,e_7\}$ and ${\bf
V}=\{e_3,e_4,e_5,e_6\}$. Now, if we reduce this along $\xi=e_{4}$,
then $V_{\xi}=< e_1,..,\hat{e}_{4},..,e_7>$ and the induced
symplectic form is $\omega_{\xi}= -e^{15}+e^{26}-e^{37}$, and the
induced complex structure is
\[
J_{\xi} =\left(
\begin{array}{ccc}
{\bf e_1}  & \mapsto  &e_5   \\
{\bf e_2} & \mapsto  &-e_6  \\
{\bf e_3} & \mapsto & e_7
\end{array}
\right)
\]
\vspace{.05in} \noindent and the complex valued $(3,0)$ form is
$\Omega _{\xi }=(e^1-ie^5)\wedge (e^2+ie^6)\wedge(e^3-ie^7)$.
\vspace{.1in}

On the other hand, if we choose $\xi '=e_{7}$ then $V_{\xi '}=<
e_1,..,e_6>$ and the symplectic form  is $\omega_{\xi '}=
e^{12}+e^{34}+e^{56}$ and the complex structure is
\[
J_{\xi '} =\left(
\begin{array}{ccc}
{\bf e_1 } & \mapsto  &- {\bf e_2 }  \\
e_3 & \mapsto  & -e_4  \\
  e_5 & \mapsto & -e_6
\end{array}
\right)
\]
\vspace{.05in} \noindent Also $\Omega_{\xi '} =(e^1+ie^2)\wedge
(e^3+ie^4)\wedge(e^5+ie^6)$.
\vspace{.1in}

In the expressions of $J$'s the basis of associative bundle ${\bf
E}$ is indicated by bold face letters to indicate the different
complex structures on $\mathbb{T}^6$. If we choose $\xi$ from the
coassociative bundle ${\bf V}$ we get the complex structure which
decomposes the 6-torus as $\mathbb{T}^3\times \mathbb{T}^3$. On the
other hand if we choose $\xi$ from the associative bundle ${\bf E}$
then the induced complex structure on the $6$-torus corresponds to
the decomposition as $\mathbb{T}^2\times \mathbb{T}^4$.

\vspace{.1in}

{\bf Remark} Notice that we are finding a mirror pair for which the
complex structures are induced from the same calibration 3-form
$\varphi$ in a $G_2$ manifold (with a technical condition that the
vector fields $\xi$ and $\xi'$ be deformations of each other). In
this sense they are dual to each other. This torus example is of
course trivial and it is well-known that $h^{(1,1)}=h^{(2,1)}=9$ for
$\mathbb{T}^6$. But in the next section we will observe that
something analogous happens in other compact $G_2$ manifolds
constructed by Joyce.

\vspace{.1in}

Next, we will show that one can write a map $H^{1,1}(X_{\beta} ) \to
H^{2,1}(X_{\alpha})$, where $\{ \alpha, \beta \}$ are orthonormal
vector fields on the $G_2$ 7-torus which give underlying ${\mathbb
T}^3\times {\mathbb T}^3$ and ${\mathbb T}^2\times {\mathbb T}^4$
Calabi-Yau decomposition, respectively. Let also $\Omega^{2,1}(TX_{
\alpha})$ and $\Omega^{1,1}(TX_{ \beta})$ be the $(2,1)$ and $(1,1)$
forms on the 3-tori $X_{ \alpha}, X_{ \beta}$, which are generated
by complex coordinates $dz_i\wedge dz_j \wedge d\overline{z}_k$ and
$dw_i\wedge d\overline{w}_j$. By using Proposition 6 in
\cite{Salur}, one can construct a natural correspondence between
$\Omega^{1,1}(TX_{ \beta})$ and $\Omega^{2,1}(TX_{ \alpha})$.

\vspace{.1in}

 Let $w_i$ be complex coordinates on ${\mathbb
T}^2\times {\mathbb T}^4$ and $z_i$ be complex coordinates on
${\mathbb T}^3\times {\mathbb T}^3$. Using the complex structures on
${\mathbb T}^3\times {\mathbb T}^3$ and ${\mathbb T}^2\times
{\mathbb T}^4$ we can write $dz_i$ and $dw_i$ in terms of the local
coordinates as follows:
$$dz_1=dx_1-idx_5, \;\;\; dz_2=dx_2+idx_6, \;\;\; dz_3=dx_3-idx_7.$$
$$dw_1=dx_1+idx_2, \;\;\; dw_2=dx_3+idx_4, \;\;\; dw_3=dx_5+idx_6.$$

\vspace{.1in}

 One can see that Re$(dz_i\wedge dz_j\wedge
d\overline{z}_k)$ and Im$(dz_i\wedge dz_j\wedge d\overline{z}_k)$ of
${\mathbb T}^3\times {\mathbb T}^3$ can be written in terms of its
$Re\;\Omega$ and $Im\;\Omega$ as follows:
$$ Re(dz_i\wedge dz_j\wedge
d\overline{z}_k)=(\partial/\partial x)\lrcorner(Re\;\Omega)\wedge
e^a + (\partial/\partial y)\lrcorner(Re\;\Omega)\wedge e^b .$$

$$ Im(dz_i\wedge dz_j\wedge
d\overline{z}_k)=(\partial/\partial x)\lrcorner(Im\;\Omega)\wedge
e^a + (\partial/\partial y)\lrcorner(Im\;\Omega)\wedge e^b .$$

\vspace{.1in}

 \noindent Here $e^a + i e^b=dz_k$ and $dx + i dy=dz_l$
for the complex coordinate $z_l$, where $i\neq l $ and $j\neq l $.
$dz_k= da + i db$, and let $dz_l= dx+ i dy $  with $l\neq i $, $ j $

\vspace{.1in}

Then by Proposition 6, in\cite{Salur}, on $X_{\alpha}$ the following
hold
$$ Re\;\Omega_{\alpha }=\omega_{\beta} \wedge \beta^{\#} + Re\;
\Omega_{\beta}. $$
$$ Im\;\Omega_{\alpha }=\alpha  \lrcorner \; (\star \omega_{\beta})-(\alpha
\lrcorner \; Im\; \Omega_{\beta} )\wedge \beta^{\#}.
$$
\vspace{.1in} By plugging in $\omega_{\beta} \wedge \beta^{\#} +
Re\; \Omega_{\beta}$ for $Re\;\Omega_{\alpha }$ and $\alpha
\lrcorner \; (\star \omega_{\beta})-(\alpha  \lrcorner \; Im\;
\Omega_{\beta} )\wedge \beta^{\#} $ for $ Im\;\Omega_{\alpha }$ one
can get
$$ Re(dz_i\wedge dz_j\wedge
d\overline{z_k})=(\partial/\partial x)\lrcorner(\omega_{\beta}
\wedge \beta^{\#} + Re\; \Omega_{\beta})\wedge e^a +
(\partial/\partial y)\lrcorner(\omega_{\beta} \wedge \beta^{\#} +
Re\; \Omega_{\beta})\wedge e^b . $$
$$ Im(dz_i\wedge dz_j\wedge
d\overline{z}_k)=(\partial/\partial x)\lrcorner(\alpha \lrcorner \;
(\star \omega_{\beta})-(\alpha  \lrcorner \; Im\; \Omega_{\beta}
)\wedge \beta^{\#})\wedge e^a$$ $$\;\;\; \;\;\;\;\;\; \;\;\;\;\;\;
\;\;\;\;\;\; \;\;\;\;\;\; \;\;\;\;\;\; \;\;\;\;\;\; \;\;\;\;\;\;
\;\;\;+ (\partial/\partial y)\lrcorner(\alpha \lrcorner \; (\star
\omega_{\beta})-(\alpha \lrcorner \; Im\; \Omega_{\beta} )\wedge
\beta^{\#})\wedge e^b ,
$$
\vspace{.1in} \noindent and writing $\omega_{\beta}$,
$Re\;\Omega_{\beta}$ and $Im\;\Omega_{\beta}$ in terms of
$dw_i\wedge d\overline{w}_j$ gives the required correspondence.

\section{Dualities related to $G_2$ manifolds}

\label{msa4}

 In a previous section we described certain compactifications
related to CY-manifolds with zero Euler number (see below
(\ref{papaton})). These spaces are special because they are the ones
which can be ``dual'' to $G_2$ holonomy spaces in the
M-theory/heterotic sense. We should be more precise in this point.
Compactification of D=11 supergravity over $G_2$ holonomy manifolds
gives abelian gauge fields and non chiral matter. Instead
compactification of heterotic string over CY spaces give chiral 4
dimensional supergravity with \emph{non}- abelian gauge fields.
Hence these compactifications can not be equivalent in general. But
if the Euler number of the CY is zero, non chiral matter is obtained
from the heterotic side. Moreover, there is an anomaly free
condition which states that Tr $R^2$-Tr $F^2$ is cohomologous to
zero and this broke the gauge group to a subgroup. These subgroup
can be further reduced by Wilson lines. It is reasonable to suppose
that, by choosing a suitable CY, this group can be reduced to an
abelian subgroup. The field strength $F$ would take values in a
abelian subgroup of $E_8 \times E_8$ or $SO(32)$, so it can be
assumed that it is $U(1)^{16}$. Under so limited circumstances, both
compactifications could be equivalent. In the abelian case, the
resulting theory from CY compactifications is $D=4$ N=1 supergravity
coupled to 16 N=1 vector multiplets and ($h_{1,1}+h_{2,1}+1$) $N=1$
scalar multiplets. Compactification of eleven dimensional
supergravity over the $G_2$ manifold gives $b_2$ vector multiplets
and $b_3$ scalar multiplets, being $b_2$ and $b_3$ the second and
the third Betti number of the $G_2$ holonomy manifold. The
equivalence can exist only if \be\lb{hip} b_2=16, \qquad
b_3=h_{1,1}+h_{2,1}+1. \ee Fortunately, there exist spaces realizing
the condition (\ref{hip}). For instance, for threefolds satisfying
(\ref{papaton}) it is deduced from (\ref{hip}) that $b_2=16$ and
$b_3=39$, and $G_2$ holonomy spaces with these Betti numbers do
exist \cite{Joyce1}. Nevertheless, even in the zero cohomology,
there are obstructions to be anomaly free, so this equivalence can
fail. This can be avoided by further compactifying to $D=3$ it is
obtained a new equivalence condition \be\lb{hung}
b_2+b_3=h_{1,1}+h_{2,1}+17, \ee which is invariant under the mirror
transformation (\ref{mirrsim}). In particular, for CY-manifolds
satisfying (\ref{papaton}), there also exist several compact $G_2$
for which (\ref{hung}) is satisfied.
\vspace{.1in}

   Other interesting dualities relating $G_2$ manifolds
are those related to II compactifications. Let us recall that for
compact $G_2$ manifold the two independent Betti numbers are $b_2$
and $b_3$, the other ones are related to them by the relations
$b_5=b_2$, $b_4=b_3$ and $b_1=b_7=1$. If IIA supergravity is
compactified over any of such manifolds the result will be $b_2+1$
vector multiplets and $b_3$ scalar multiplets. For IIB
compactifications the result is $b_2+1$ scalar multiplets and $b_3$
vector multiplets. Naively, this indicate that the Betti numbers of
mirror $G_2$ manifolds are related by $b_2+1\leftrightarrow b_3$.
But in three dimensions scalars and vector multiplets are related by
a duality transformation. Thus this result only implies that
manifolds with the same $b_2+b_3$ are collected together. This fact
was noticed already in \cite{ShataVafa} and \cite{Joyce1}. Also in
\cite{Acharya2} there analyzing the spectrum of both theories in
orbifolds of the form ${\mathbb T}^7/{\mathbb Z}_2\oplus {\mathbb
Z}_2\oplus {\mathbb Z}_2$. It was found that T-duality interchange
IIA/IIB theories with (without) discrete torsion to IIB/IIA without
(with) discrete torsion. In many cases, the Betti numbers are inert.
But also there were found certain examples with different Betti
numbers giving rise to (apparently) the same compactification
\cite{Acharya2}.
\vspace{.1in}

   The special examples in which the Betti numbers stands unchanged
under mirror symmetry correspond to the so called "Joyce manifolds
of the first kind", which were the only examples known when the
previous analysis was done. It could be interesting to understand
the geometric property which "protects" them under the mirror
transformation. The formalism presented in the previous section and
a description of the Joyce space will be very important for this
purpose.

\subsection{Joyce $G_2$ manifolds of first kind}

 These are compact 7-manifolds with holonomy $G_2$.
An intuitive idea of their construction comes from compactifications
of eleven dimensional SUGRA to $d=4$ with orbifolds of ${\mathbb
T}^7$ as internal spaces. When eleven dimensional supergravity is
compactified over ${\mathbb T}^7$ the number of supersymmetries of
the four dimensional theory is $N=8$. A ${\mathbb Z}_2$ action will
kill two supersymmetries and we will have $N=4$. A further quotient
by a new ${\mathbb Z}_2$ commuting with the first one will kill
again half of the supersymmetries, so the number will be $N=2$.
Still another ${\mathbb Z}_2$ action will reduce the supersymmetry
to $N=1$, which is the number of supersymmetries obtained from
compact $G_2$ manifolds. This suggest the possibility of
constructing compact $G_2$ manifolds by starting with a quotient of
${\mathbb T}^7$ by a suitable ${\mathbb Z}_2\oplus {\mathbb
Z}_2\oplus {\mathbb Z}_2$ group, and blowing up conveniently the
singular set. This ideas has certain analogy with the $K3$ case.
\vspace{.1in}

   In fact, this idea was implemented effectively in
\cite{Joyce1}. A group ${\mathbb Z}_2\oplus {\mathbb Z}_2\oplus
{\mathbb Z}_2$ affecting the seven coordinates of the torus is
$\Gamma=\{\alpha, \beta, \gamma\}$, being
$$
\alpha(x_1,\cdots,x_7)=(-x_1,-x_2,-x_3,-x_4,x_5,x_6,x_7),
$$
\be\lb{goma}
\beta(x_1,\cdots,x_7)=(b_1-x_1,b_2-x_2,x_3,x_4,-x_5,-x_6,x_7), \ee
$$
\gamma(x_1,\cdots,x_7)=(c_1-x_1,x_2,c_3-x_3,x_4,c_5-x_5,x_6,-x_7),
$$
\vspace{.1in}
\noindent

with $b_1$, $b_2$, $c_1$, $c_3$, and $c_5$ some constants in the
interval $\{0,\frac{1}{2}\}$. It is not difficult to check that the
transformations $\alpha$, $\beta$ and $\gamma$ are ${\mathbb Z}_2$
actions and commute thus, one can blowup their singular set
successively \cite{Liu}. Consider first ${\mathbb T}^7/\alpha$. An
inspection of (\ref{goma}) shows that $\alpha$ affect the first four
coordinates of ${\mathbb T}^7$ and leave the other three inert. This
means that resolving the singular set of ${\mathbb T}^7/\{\alpha\}$
is equivalent to resolving ${\mathbb T}^4/\{\alpha\}\times T^3$,
where ${\mathbb T}^3$ is the torus parameterized by the coordinates
$x_5$, $x_6$ and $x_7$. By replacing balls $B^4/{\mathbb Z}_2$
around the singularities by copies of the Eguchi-Hanson space and
making a blowup we will obtain $K3\times {\mathbb T}^3$ as a
resolution.
\vspace{.1in}

  The next task is to find the effect of $\beta$ on $K3\times {\mathbb
T}^3$, which will follow from the effect on ${\mathbb
T}^4/\{\alpha\}\times T^3$. From (\ref{goma}) it is seen that the
coordinate $x_7$ is rigid under the action of $\beta$. Let us
parameterize the remaining 6-torus ${\mathbb T}^6$ with the
coordinates $z_1=x_1+ix_2$, $z_2=x_3+ix_4$, $z_3=x_5+ix_6$ and their
complex conjugates $\overline{z}_i$. The K\"{a}hler form of
${\mathbb T}^6$ is $\overline{J}_6=dz_i\wedge d\overline{z}_i$ and
$\beta^{\ast}(\overline{J}_6)=\overline{J}_6$. Thus $\beta$ acts
holomorphically on such torus. In similar fashion, $\gamma$ acts
anti-holomorphically. Let us decompose ${\mathbb T}^6={\mathbb
T}^4\times {\mathbb T}^2$, where the 4-torus is parameterized by
($z_1$, $\overline{z}_1$, $z_2$, $\overline{z}_2$) and the 2-torus
is parameterized by ($z_3$, $\overline{z}_3$). Then the
$\beta$-action on ${\mathbb T}^2$ is $(-1)$. Also, the $\beta$
action over the K\"{a}hler two form $\overline{J}_1+i
\overline{J}_2=dz_1\wedge dz_2$ of ${\mathbb T}^4$ is $(-1)$. After
resolving the singular set of $\alpha$, the torus ${\mathbb T}^4$
becomes $K3$ and the two-form $dz_1\wedge dz_2$ descends to a
holomorphic two form $\overline{J}$ over K3. Then the $\beta$ action
over ${\mathbb T}^4$ descends to an action $\sigma$ on $K3$ that
acts on $\overline{J}$ as $\sigma(\overline{J})=-\overline{J}$. In
conclusion, $\beta$ descends on $K3\times {\mathbb T}^2$ to the
automorphism $(\sigma, -1)$ as for the Borcea-Voisin case
(\ref{borcea}). The corresponding elliptic curve is $E={\mathbb
T}^2$. Thus the resolution of ${\mathbb T}^7/\{\alpha,\beta\}$ is a
trivial bundle $X_6\times {\mathbb T}^1_7$, where $X_6$ is the
resolution of (\ref{borcea}) with $E={\mathbb T}^2$. \vspace{.1in}

    The final step is to find the effect of $\gamma$ over
$X_6\times {\mathbb T}^1_7$. As we have seen, $\gamma$ is
anti-holomorphic on the K\"{a}hler form $\overline{J}_6$ of
${\mathbb T}^6$ and descends to an anti-holomorphic action over
$X_6$. The calibration form $\varphi$ of $X_6\times {\mathbb T}^1_7$
is
$$
\varphi=dx_7\wedge \overline{J}_6^{\prime}+Re (\Omega^{\prime})
$$
$$
\ast\varphi=dx_7\wedge (Im
\Omega^{\prime})+\frac{1}{2}\overline{J}_6^{\prime}\wedge
\overline{J}_6^{\prime}
$$
\noindent where $\overline{J}_6^{\prime}$ is the K\"{a}hler form of
the Borcea-Voisin threefold and $\Omega^{\prime}$ its complex closed
three form. As $\gamma$ acts as a negation on $x_7$ we conclude from
the $\gamma$-invariance of $\varphi$ and $\ast\varphi$ that
$\gamma(\overline{J}_6^{\prime})=-\overline{J}_6^{\prime}$ and that
$\gamma(\Omega^{\prime})=\overline{\Omega}^{\prime}$. Therefore
$\gamma$ descends to an anti-holomorphic action on the CY threefold
and as a negation on $S^1_7$. Furthermore, from (\ref{goma}) it
follows that the fibers corresponding to $x_7=0$ and
$x_7=\frac{1}{2}$ are $\gamma$-invariant. The singular set of
$X_6\times {\mathbb T}^1_7$ is located in these fibers and consists
on the $3$-tori or the free ${\mathbb Z}_2$-quotient of $3$-tori
descending from the fixed ${\mathbb T}^3$'s elements of $\gamma$,
$\alpha\gamma$, $\beta\gamma$, or $\alpha\beta\gamma$ on ${\mathbb
T}^7$. The neighbor of the singularities is of the form ${\mathbb
T}^3\times B^4/{\mathbb Z}_2$ and the resolution is by blowups as in
the Kummer case. The result are the Joyce spaces of the first kind,
which therefore are fibrations over Borcea-Voisin CY three folds
\cite{Liu}.

\subsection{A mirror pair inside a $G_2$ space}

  We have seen in a previous section that $\mathbb{T}^7$ with its flat
$G_2$ structure $\varphi$ induces a mirror pair of CY´s, which is
the torus $\mathbb{T}^6$ decomposed as $\mathbb{T}^4\times
\mathbb{T}^2$ and as $\mathbb{T}^3\times \mathbb{T}^3$. This provide
a mapping between the complex and sympletic structure required by
mirror symmetry and was obtained by reduction along an associative
and coassociative cycle respectively. This example is rather simple,
because $\mathbb{T}^6$ is mirror to itself.
\vspace{.1in}

  Nevertheless, we can show that
something analogous happens for Joyce manifolds of the first kind,
this issue has also been analyzed in \cite{akbulut}. As they are
locally fibrations over Borcea-Voisin 3-folds obtained by resolving
the singularities of orbifolds of $\mathbb{T}^6$, it is plausible to
guess that the CY base spaces have $(h_{1,1},h_{2,1})=(19,19)$. Note
that if this were the case, the base of the fibration is protected
again the complex/sympletic mapping, that is, this mapping does not
change the topology of the manifold.
\vspace{.1in}

   We can directly check that our previous discussion is true.
The 3-folds in consideration arise as resolutions of $\mathbb{T}^6$
divided by the $\alpha$ and $\beta$ actions of (\ref{goma}), which
are commuting $\mathbb{Z}_2$ actions. The first ${\mathbb Z}_2$
generator possess a singular set consisting in sixteen copies of
${\mathbb T}^2$. But these copies are interchanged by the second
generator, leaving eight invariant torus. The same argument is true
for the second generator so the total number of ${\mathbb T}^2$
copies is 16. For ${\mathbb T}^6$ we have $(h_{1,1},h_{2,1})=(3,3)$
and any of these ${\mathbb T}^2$ copies add 1 to the Hodge numbers.
The result is finally $(h_{1,1},h_{2,1})=(19,19)$. Thus, the
Borcea-Voisin manifolds presented in the previous section are of
zero Euler number and satisfy the constraint (\ref{papaton}).

\vspace{.1in}

    In fact, we can make another computation of the Hodge numbers
(\ref{borcea}) giving the same result. From (\ref{mu}) we see that
the calculation of the Hodge numbers is related to find the
components of the singular set $\Sigma$ of the $\sigma$ of
(\ref{borcea}), which acts over the K3 surface. An inspection of
(\ref{goma}) shows that the $\beta$ action over ${\mathbb T}^4$ has
4-fixed points, but they are interchanged by the $\alpha$ action and
thus there are essentially two components. This means that
$n^{\prime}=2$ in (\ref{mu}). Also, in ${\mathbb T}^6$ the
neighborhood of the singularities is of the form ${\mathbb T}^2$,
and these tori are disjoint. It means that $\Sigma= {\mathbb
T}^2\cup {\mathbb T}^{2\prime}$. Thus $n=2$ and we obtain from
(\ref{mu}) that
$$
h^{1,1}(Y)=h^{2,1}(Y)=19.
$$
So, we have checked that the calculation is correct. Thus, we have
found a mirror CY structure inside Joyce manifolds of the first
kind. Note that this result does not depend on the election of the
constants $b_i$ and $c_i$ of the actions (\ref{goma}).

\vspace{.1in}

  Let us go back to our original motivation, which is to understand why
"apparently" mirror symmetry leave the betti numbers of the Joyce
spaces inert. As the underlying CY base is protected under the
complex/sympletic mapping, if the fibration were trivial, the same
will occur with the entire 7-manifold. We suggest that the situation
still holds after performing the quotient by $\gamma$ and perturbing
the metric around the singularities. This is an plausibility
argument only. We also suggest that this situation is generalized to
barely $G_2$ manifolds, which we describe next.

\vspace{.1in}

\subsection{Joyce spaces as ``barely'' $G_2$ manifolds}

Summing up our discussion at this point, Joyce spaces of the first
kind arise as resolutions of the singular set of
$$
X_7=\frac{(X_6 \times S^1)}{{\mathbb Z}_2}
$$
being the ${\mathbb Z}_2$ action identified with $\gamma$ and $X_6$
the CY 3-fold described in (\ref{borcea}). $\gamma$ is an
anti-holomorphic action on the $X_6$ threefold and as a negation on
$x_7$. Such kind of $G_2$ manifolds are called ``barely'' $G_2$
manifolds. In other words barely $G_2$ manifolds arise as
resolutions of orbifolds of the form \be\lb{bare} X_7=\frac{(CY
\times S^1)}{{\mathbb Z}_2}, \ee being CY a Calabi-Yau manifold. The
action ${\mathbb Z}_2$ is given by $(\sigma, -1)$ being $\sigma$ a
real structure, which acts as an isometry of the CY for which
$$
\sigma^{\ast}(\Omega)=\Omega, \qquad
\sigma^{\ast}(\overline{J})=-\overline{J},
$$
\noindent where $\Omega$ is the closed holomorphic three form and
$\overline{J}$ the K\"{a}hler two form.

\vspace{.1in}

  Several aspects of barely $G_2$ manifolds are already
investigated and can be applied for Joyce manifolds. The associative
three cycles of barely $G_2$ manifolds fall into two types
\cite{Moore}. The first are of the form \be\lb{first}
\Sigma_{hol}=\frac{(\Sigma_2\times S^1) }{{\mathbb Z}_2}\ee where
$\Sigma_2$ is a holomorphic cycle on the CY which is mapped to
$-\Sigma_2$ by the action of $\sigma$. If $\Sigma_2$ is rational
then $\Sigma$ is a rational 3-sphere. The other associative
submanifolds are \be\lb{second} \Sigma_r=\frac{(\Sigma^+)}{{\mathbb
Z}_2} \ee where $\Sigma^+$ is a Lagrangian 3-cycle in the CY
manifold. These are mapped to $-\Sigma^+$ by the action of $\sigma$.
\vspace{.1in}

 Also, in M-theory compactifications
on such ``barely'' backgrounds there are no 5-brane instantons
\cite{Moore}. Therefore we conclude that \emph{compactifications of
eleven dimensional supergravity over Joyce spaces of the first kind
give no 5-brane instantons}. This implies that contribution to the
superpotential is given in terms of the associative cycles, i.e,
$W=W(\Sigma_{hol})+W(\Sigma_{r})$. Other aspects of strings
propagating in barely $G_2$ spaces have been also considered for
instance, in \cite{Eguchi} and \cite{Roiban}.
\vspace{.1in}

  Finally, for barely $G_2$ manifolds we have that
$b_2+b_3=h^{1,2}+h^{1,1}+1$. Notice that this relation is invariant
under the complex/sympletic map $h^{1,2}\leftrightarrow h^{1,1}$ of
the underlying CY.

\section{Interpretation}

  Our aim is now to interpret the presented results in the context
of mirror symmetry. We suggest that the reason for which the mirror
map apparently leave the betti numbers of the Joyce spaces of the
first kind inert \cite{PapadopoulosTownsend} is that they are
fibrations over CY 3-folds which are "protected" under the action
complex/sympletic map. In other words, this map does not affect the
topology of the underlying CY base. We generalize this suggestion to
"barely" $G_2$ manifolds by supposing that \emph{for any barely
$G_2$ manifold fibered over a CY $X_6$ the natural mirror candidate
is another barely $G_2$ manifold fibered over the mirror
$X_6^{\prime}$ of $X_6$}. In fact the sum
$b_2+b_3=h^{1,2}+h^{1,1}+1$ will be the same in both cases. Although
this is not conclusive evidence we propose that, as $X_6 \times S^1$
and $X_6^{\prime}\times S^1$ are mirror 7-manifolds, the mirror
property is (approximately) preserved after dividing by a ${\mathbb
Z}_2$ action and perturbing the resulting orbifolds to smooth $G_2$
holonomy metrics.

\vspace{.1in}

    We should make some comments about
this suggestion. First, the mirror $G_2$ pairs found till the moment
are not exact. The resulting theories agree up to certain extent,
but they are not completely isomorphic \cite{Acharya2}. Second, our
suggestion should be understood as a classical one for the following
reason. The untwisted sector of II strings propagating on such
manifolds contains massless states consisting $b_2+b_3$ chiral
multiplets. But has been shown in \cite{Eguchi} the appearance of
additional massless states in the twisted sector. These states were
interpreted in terms of quantum effects \cite{Eguchi}. Classically,
if the anti-holomorphic action has no fixed points on the CY, one
does not expect new massless states to appear. Thus, our statement
make sense only as classical one.

\vspace{.1in}

    There are other compact examples of $G_2$ holonomy
manifolds found by Kovalev, \cite{Kovalev}, which are constructed by
gluing asymptotically cylindrical $G_2$ manifolds along the
boundary. It sounds reasonable to suppose that the resulting
manifold can be also described as some kind of CY fibration. This
leads to the following question:

\vspace{.1in}

{\bf Question} {\em For any given compact $G_2$ manifold, does there
exist an open subset which is of the form Calabi-Yau $\times$ (open
interval)? }

\vspace{.1in}

An affirmative answer to this question will imply that compact
mirror $G_2$ manifolds admit a description in terms of mirror CY
manifolds. In any case, we feel that a deep study of the
submanifolds inside compact $G_2$ manifolds is worthy.

\vspace{.1in}

   As a future investigation it will be interesting to
repeat the analysis performed in \cite{Eguchi} for the barely $G_2$
manifolds presented here and for those presented in
\cite{Partouche}. We also would like to make a more precise
definition of mirror symmetry for $G_2$ manifolds, perphaps using
the topological versions presented in \cite{Boer}-\cite{Boer2}. We
leave this for a future investigation.
\vspace{.1in}

{\small{\it Acknowledgements.} The first-named author would like to
thank IMBM, Istanbul Center for Mathematical Sciences for providing
an excellent research environment. Special thanks to Betul Tanbay
for her support and hospitality. The second-named author thanks to
ICTP, Trieste, where part of this work was performed and to the
Secretary of Culture of Avellaneda for support.

\end{document}